\documentstyle[twoside,fleqn,espcrc2]{article}

\input epsf.tex

\def\be{\begin{equation}}
\def\ee{\nonumber\end{equation}}
\def\bea{\begin{eqnarray*}}
\def\eea{\end{eqnarray*}}
\def\bc{\begin{center}}
\def\ec{\end{center}}

\def\IR{{\hbox{{\rm I}\kern-.2em\hbox{\rm R}}}}
\def\IB{{\hbox{{\rm I}\kern-.2em\hbox{\rm B}}}}
\def\IN{{\hbox{{\rm I}\kern-.2em\hbox{\rm N}}}}
\def\IC{{\hbox{{\rm I}\kern-.6em\hbox{\bf C}}}}
\def\IZ{{\hbox{{\rm Z}\kern-.4em\hbox{\rm Z}}}}

\def\semi{;\hfill\break}

\newcommand{\AmS}{{\protect\the\textfont2
  A\kern-.1667em\lower.5ex\hbox{M}\kern-.125emS}}

\title{Introduction to D--Branes, with Applications}

\author{Clifford V. Johnson\address{Institute for Theoretical Physics,
        University of California, Santa Barbara, CA 93106 U.S.A.}
        \thanks{email: {\tt cvj@itp.ucsb.edu}\hfill\break This work
        (report no. NSF-ITP-96-57) was supported in part by the
        National Science Foundation under grant \#PHY94--07194.}  }
       
\begin{document}

\begin{abstract}
A brief review of some of the central ideas, terminology and
techniques of the technology of orientifolds and D--branes is
presented. Some applications are reviewed, including the construction
of dual solitonic strings in the context of string/string duality, the
computation of the Bekenstien--Hawking entropy/area law for extremal
black~holes, and the construction of ${\cal N}{=}1$ string vacua in
dimensions lower than ten. \break(Presented at SUSY~'96,
University of Maryland at College Park, May 29--June 1, 1996.)
\vskip-0.5truecm
\end{abstract}

\maketitle

\section{Overview}

Studies of String Theory have traditionally been performed
perturbatively\cite{gsw}.  These studies have led to great insight
over the years into a rich theory of gravity and possibly the other
interactions of the physical world. However, many vital issues have
had to be left to await the advent of a non--perturbative
understanding of string theory.

Such an understanding has begun to emerge over the last year or
so. We are still quite a way (it~seems) from addressing  real world
problems with these new studies, but the time will surely come when
we will begin to see if we can understand more about Nature with this
approach.

Many of the exciting results from strong coupling string theory have
been obtained (or confirmed) with the package of
techniques\cite{dnotes}\ involving {\bf
`Orientifolds'}\cite{sagnotti,petrii}\ and {\bf
`D--Branes'}\cite{newcon,petri}.

Generally, D--branes are simply a more precise language with which we
can construct perturbative string backgrounds. The advantages come
mainly from the fact that certain properties of D--branes allow us to
make powerful statements about such perturbative backgrounds which
remain true beyond perturbation theory.

\section{The  Perturbative String Theories}
There are five basic superstring theories which may be perturbatively
formulated in ten dimensions\cite{gsw}:

\subsection{The ${\cal N}{=}2$ Theories}
The Type IIA and Type IIB theories describe two closed string theories
which arise from the choice of how one puts together the right and
left moving world--sheet sectors to yield the
generators of ${\cal N}{=}2$ spacetime supersymmetry.  Choice~$A$
corresponds to realising a left--moving spinor and a right--moving
spinor.  Choice~$B$ corresponds to realising spinors of the same
chirality, giving rise to a chiral theory.

\subsection{The ${\cal N}{=}1$ Theories} 
The Heterotic string theories describe closed strings with one
supersymmetry from either the left or right moving sector. The other
sector is not supersymmetric, and instead has a current algebra which
gives rise to spacetime gauge symmetry $SO(32)$ or $E_8{\times}E_8$.

The Type~I theory is a theory of open strings, with gauge symmetry
$SO(32)$.  We shall review it in some detail in the next section, and
recall how to realise it as the simplest example of an orientifold,
involving D--branes.

\section{Type I from Type IIB: An Orientifold}

{Recall that for the type IIB theory, there was a symmetry between
left and right moving structures on the world sheet. Let us call the
generator of this symmetry $\Omega$. It is simply world sheet parity,
exchanging left with right.}

{We can try to construct a new string theory by gauging this symmetry:
Project out all states in the theory which are not invariant under
$\Omega$.} This results in a new closed string theory which is
unorientable (the left and right are now indistinguishable) and so
the world sheets for string perturbation theory are unorientable.  It
is also an ${\cal N}{=}1$ supersymmetric theory,  the supersymmetry
generator being a linear combination of the two which we started with.

Unfortunately, it is inconsistent. For example, at one loop there is a
divergence in the Klein bottle diagram:

\medskip

\centerline{\epsfxsize=1.5in\epsfbox{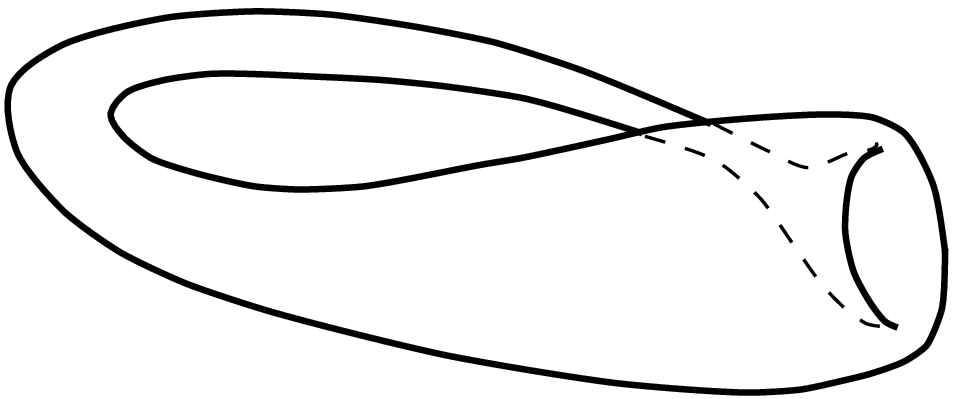}}

\medskip

Meanwhile, one might try to construct a purely open ${\cal N}{=}1$
superstring theory, endowing the string endpoints with charges ({\bf
`Chan--Paton factors'}) in the fundamental of some group $G$.
This open string theory has spacetime gauge symmetry group $G$. (A
massless spin one excitation with values in the adjoint of $G$
appears.)

Unfortunately, it is also divergent at one loop where the string
theory tries to emit massless closed string states. There are two such
divergent one loop diagrams:

\medskip

\centerline{\epsfxsize=2.5in\epsfbox{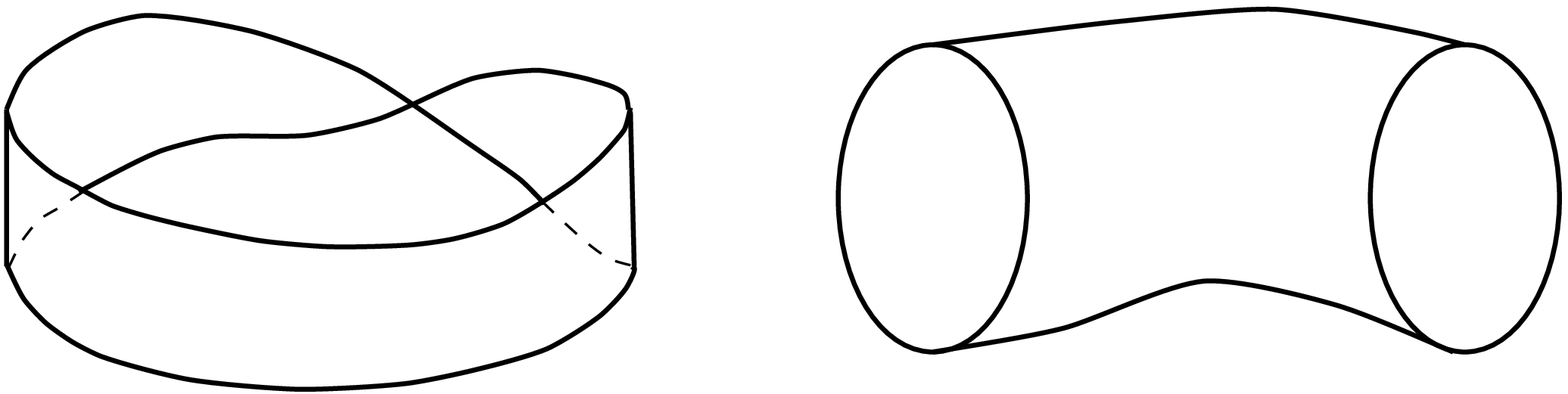}}

\medskip

If we realise that the closed string sector appearing at one loop for
this open string theory and the closed string we got from gauging
$\Omega$ in the IIB theory are to be identified, we have a chance of
cancelling these divergences against one another, leaving a consistent
theory.  The three divergent diagrams which occur are the M\"obius
strip, Klein bottle and cylinder:

\medskip

\centerline{\epsfxsize=3.0in\epsfbox{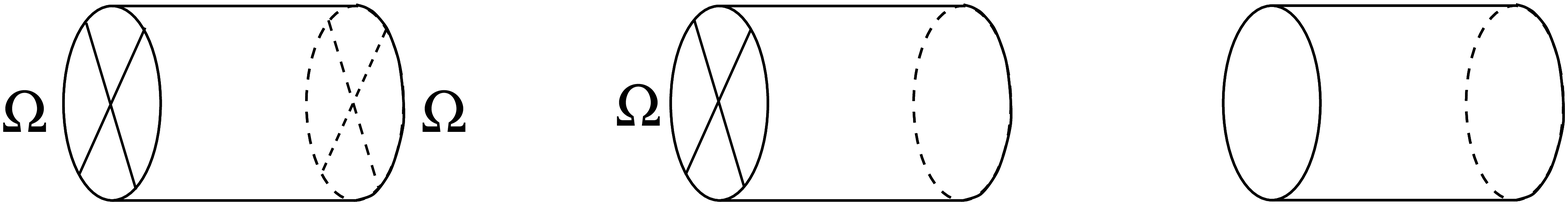}}
\medskip

It turns out that the divergences are given by a piece which is
common to all of the amplitudes, times $32^2$ for the Klein bottle,
$\pm2n_9\times 32$ for the M\"obius strip and $n_9^2$ for the
cylinder, for Chan--Paton factors corresponding to a gauge group
$SO(n_9)$ for the $-$~sign, and $U\!Sp(n_9)$ for the $+$~sign.
This can be written as $(n_9\pm32)^2$, which cancels to zero for
gauge group $SO(32)$.

We have thus recovered another chiral ${\cal N}{=}1$ spacetime
supersymmetric string theory, with gauge group $SO(32)$, this time an
open string theory.

We shall soon reinterpret this computation in the language of D--branes.

\subsection{Massless fields}

In types IIA and IIB string theory, there are the usual massless
fields, $G,\Phi$ and $B_2$, (graviton, dilaton and antisymmetric
2--form tensor) of a relativistic closed string, coming from the
Neveu--Schwarz-Neveu--Schwarz (NS-NS) sector.

It is sometimes useful to think of the fundamental IIA or IIB string
as carrying electric $H_3$ charge, where $H_3$ is the 3--form field
strength of the potential $B_2$.  The two dimensional worldsheet of
the string has a natural coupling to the 2--form potential
$B_2$. (Analogous to the familiar case of a charged particle coupling
naturally to the 1--form Maxwell potential of electromagnetism via
its world--line.)  

There are also antisymmetric $p$--form fields $A_p$, coming from the
Ramond-Ramond (R-R) sector.  For type~IIA there are $ A_1, A_3, A_5,
A_7$ and $A_9$ and for type~IIB, $A_0, A_2, A_4, A_6$ and
$A_8$.

\section{$p$--Branes}

Imagine an extended object in the theory with $p$~spatial
dimensions. ($p{=}1$ for a string--like object).   Imagine further
that such an object arises if there is a $(p{+}1)$--form in the theory
to couple naturally to its $(p{+}1)$--dimensional world volume.

Such  objects are called  {\bf `p--branes'}\cite{duff}.

So $p$--branes with $p$ odd will arise in~IIB string theory while for
$p$ even, they arise in Type~IIA theory.  They have (electric) charge
$\mu_p$ with respect to the field strength $F_{p+2}$. Ten~dimensional
Hodge duality tells us that a $p$--Brane also carries magnetic charge
with respect to a form $F_{8-2}$ and is therefore dual to a
$(6{-}p)$--Brane.

A generalisation of Dirac's argument\cite{general}\ for electric and magnetic
charges must apply here, setting the fundamental units of electric and
magnetic charge.

\section{D--Branes}

It transpires\cite{gojoe}\ that there is a special class of
$p$--branes which {\it carry the most basic unit of $F_{p+2}$ charge
allowed:} they are called {\bf `D--branes'}.  We may think of D--branes
as the consequence of considering more general boundary conditions in
string theory. The world--volume of the D--brane is the sub--manifold
in spacetime upon which strings may have end--points. Specifically, the
`D$p$--brane' is defined by  Neumann boundary conditions in
$(p{+}1)$ directions (the world--volume) and Dirichlet in the
remaining transverse directions.

The pure open string theory we saw earlier is a very special example
of a D--brane configuration: Spacetime itself contains D9--branes
(with a ten dimensional world--volume).  The (Chan--Paton) internal
degree of freedom is the choice that a string endpoint has about which
D9--brane it can end on. Gauge group $SO(32)$ corresponds to the
presence of 32 D9--branes.

Think of the diagrams above as cylinders with either open string
boundaries or crosscaps at either end.  The boundary is a D9--brane.

\section{The Divergences}

The divergences come from the regions of moduli space where the
cylinders' lengths go to infinity.  In this limit, the dominant
contribution is from massless states in the closed string sector
propagating.

We may think\cite{polcai,gojoe}\ of the above cancellation as being between
{\bf tadpoles} coming from the emission of massless closed string
states from a D9--brane (via the boundary term), or from the vacuum
(via the crosscap).

\medskip

\centerline{\epsfxsize=1.5in\epsfbox{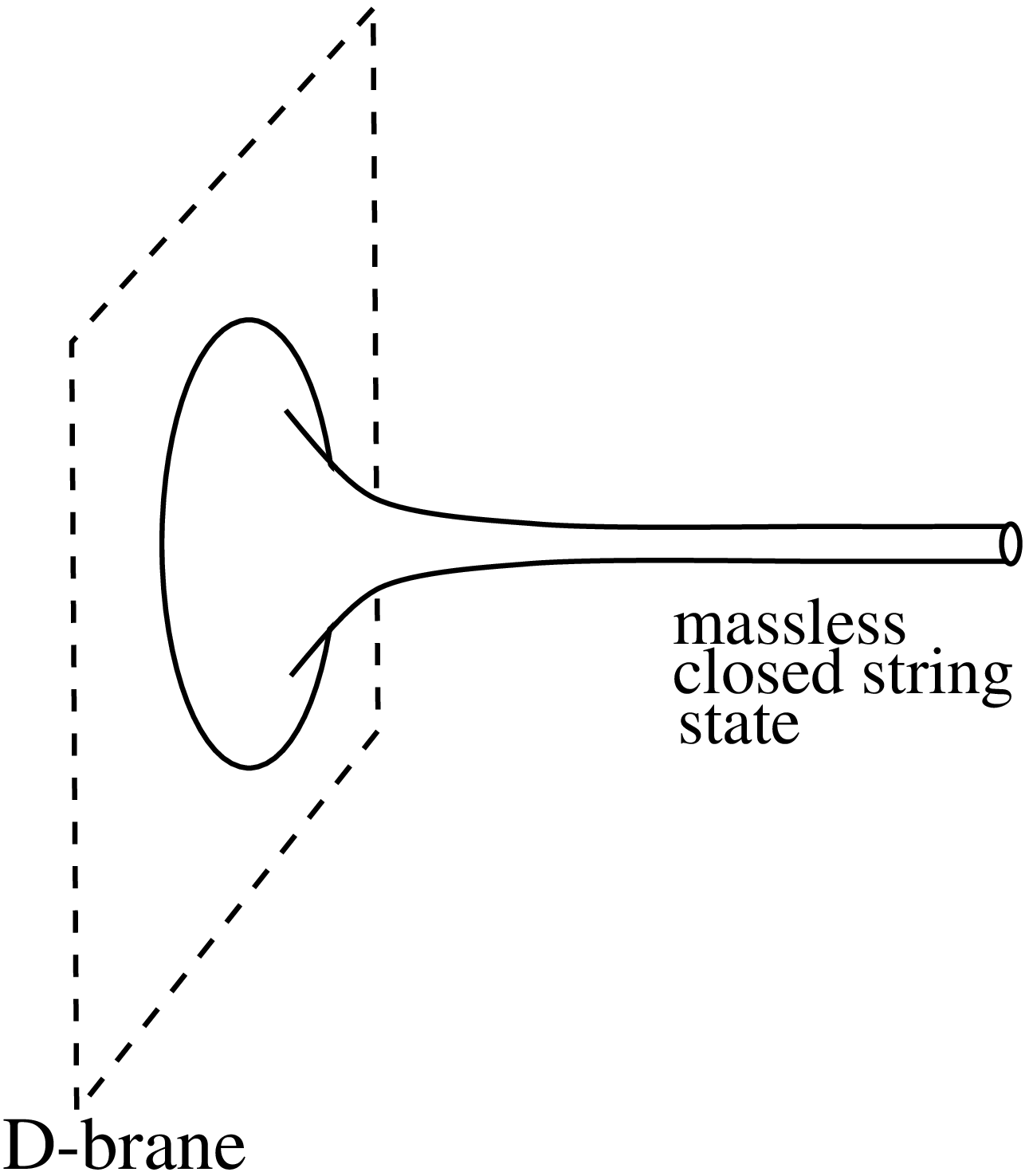}}

\medskip

This cancellation is associated with D9--branes. It is a requirement
for consistency of the field equations of the 10--form R-R
potential\cite{polcai,curt}.

\section{D--Branes and $T$--Duality}

Another useful way to see the D--branes quite clearly is to imagine
that the tenth dimension $X^9$ is a circle of radius $R_9$. {\bf
`$T$--duality'} along $X^9$ gives an open string theory compactified
on a circle of radius $R^\prime{=}\alpha^\prime/R_9$ which describes
the same physics in new variables\cite{kikk}. $T$--duality is a
symmetry which can be described perturbatively as an action on the
world sheet fields of the string theory\cite{newcon}. One of its
operations is to exchange Dirichlet with Neumann boundary conditions
along the $X_9$ direction. This turns the 32 D9--branes into 32
D8--branes.  Their world--volume is along the $\{X^0,\ldots,X^8\}$
directions. Also, the action of $\Omega$ gets converted to a manifest
spacetime reflection symmetry.  The D8--branes have a transverse
position coordinate on the $X^9$ circle (below, a line denotes the
whole world-volume; a dotted line is an orientifold plane about which
there is a reflection):

\bigskip

\centerline{\epsfxsize=3.0in\epsfbox{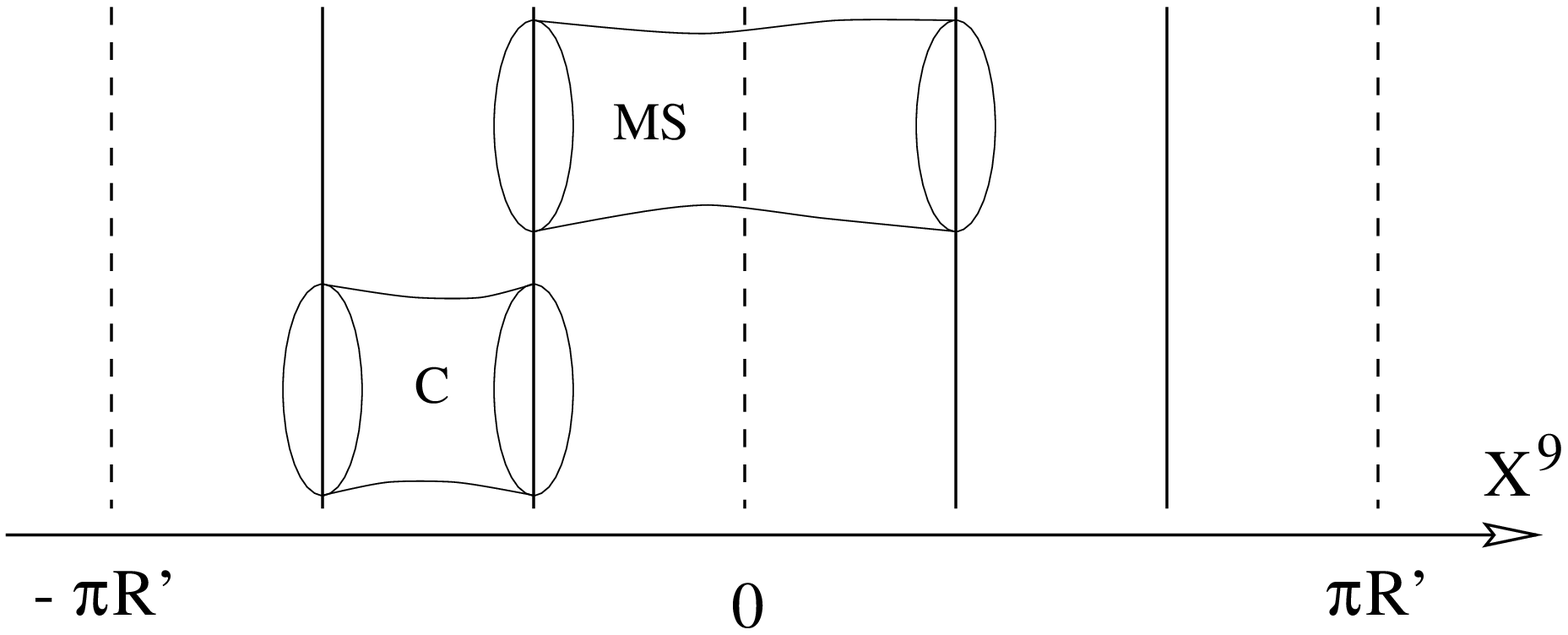}}

\medskip

The one--loop diagrams seen earlier can be seen to arise clearly here
as fundamental strings stretching between different branes. In the
figure, the cylinder (C) is formed of strings stretching between
distinct branes, while the M\"obius strip (MS) comes from strings
stretching between a brane and its mirror forming a one--loop diagram
with a single boundary.

It also can be seen here is that $SO(32)$ is an enhanced gauge
symmetry which occurs when the 32 branes and their images are all
coincident. Moving them away from each other is equivalent to
introducing Wilson lines in the dual picture to break some of the
gauge symmetry. A single isolated D--brane has gauge symmetry $U(1)$.
Enhanced gauge symmetry arises when extra modes become massless,
originating from strings stretched between coincident D--branes.

\section{D--Branes as BPS States}

An important class of states in supersymmetric theories are those
which saturate a Bogomol'nyi--Prasad--Sommerfield\cite{bps}\ ({\bf
BPS}) bound.  This is a lower bound on the mass of a state with
respect to the central charges of the supersymmetry algebra.  Well
known examples of such states are furnished by the soliton sector of
supersymmetric field theories.

There are a number of interesting and important properties of BPS
states (see ref.\cite{duff}\ for more discussion and references):

\begin{itemize}
\item{They break half the supersymmetry: Half of the supercharges
annihilate them, the other half don't.}
\item{The force between such BPS states is zero.}
\item{The spectrum of masses and charges of BPS states in a given
theory is exact: It may be computed at weak coupling, and then there
is a non--renormalisation theorem which protects this spectrum from
corrections, to all orders in perturbation theory.}
\end{itemize}

{\it D--Branes are BPS states\cite{gojoe}}.  They break half the
supersymmetries, have zero force between each other, and their
masses\cite{newcon}\ (${\sim}1/g_{\rm str}$) saturate the bound.

\section{Application I:  String/String Duality in Ten Dimensions}

In the field theory examples of strong--weak coupling duality, one of
the basic phenomena is the exchange of roles of the light,
`fundamental' charge carriers (electrons) with the heavy `solitonic'
dual degrees of freedom (monopoles), as the coupling goes large.

In a supersymmetric setting, these solitons will be BPS states whose
properties we can study, trusting that these properties will survive
at strong coupling.

Many of the strong--weak coupling duality examples studied in the last
year have urged us to identify the soliton in the theory which becomes
light at strong coupling. It will carry the fundamental degrees of
freedom in the dual theory.

In conjecturing strong--weak coupling duality relations between string
theories, we can test the conjecture by studying the properties of the
candidate soliton string which will become light at strong coupling.
As each of the string theories has very distinct signature properties,
it is easy to see if the conjecture has a good chance of being right,
if one has the tools to look for these properties\cite{duals}.

D--branes are the appropriate tools in many of these examples.

\subsection{Type IIB Self Duality and $SO(32)$ Type I/Heterotic Duality}

Two examples of a conjectured string--string duality in ten dimensions
are the self duality of the type IIB string\cite{hull}, and the
duality relation between the $SO(32)$ string theories\cite{ed}.

The obvious candidate soliton string which will become light at strong
coupling in each theory is the D1--brane.  As a soliton, it is very
heavy at weak string coupling.  However, its mass is inversely
proportional to the string coupling, a basic property of
D--branes\cite{newcon}, and so it will become light at strong
coupling.

In both cases it has to have the right
properties to become the dual string in the strong coupling limit:
One can use basic D--brane calculus to study their
world--sheet properties.  The excitations of a D--brane can be studied
by examining the fundamental strings which end on the brane, defining
its position and dynamics:

\centerline{\epsfxsize=1.5in\epsfbox{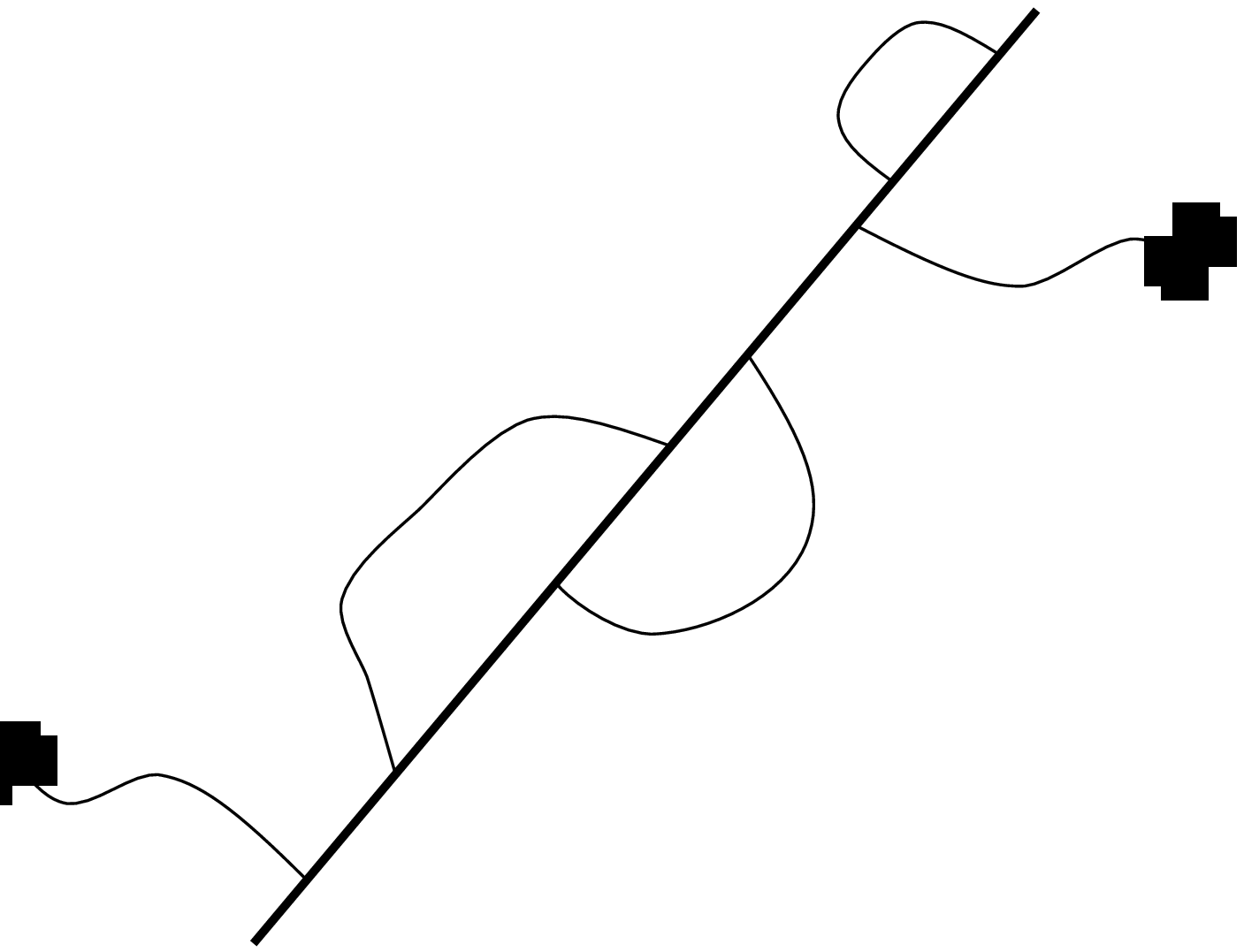}}

Embedding such a macroscopic string in the background produces a
decomposition of the spacetime Lorentz group $SO(1,9)$ to
$SO(8){\times}SO(1,1)$. (The factors refer to directions transverse
and parallel to the world--volume, respectively.)

The D1--brane in the type IIB theory  has been shown to have right and
left moving (with respect to $SO(1,1)$) fermions on its world--sheet
which are of the same chirality $SO(8)$ spinors: the massless
fermionic content of a IIB superstring\cite{edbound}.

Meanwhile, after projecting with $\Omega$ and adding D9--branes to get
the type I theory, the D1--brane in that theory has only one $SO(8)$
spinor, coming from (say) the left, while the 1-9 and 9-1 fundamental strings
produce a right--moving $SO(32)$ current algebra. This is the world
sheet content of a heterotic string\cite{edjoe}.

\section{Application II: Extremal Black Hole Entropy}

Another important application has been the demonstration that the
Bekenstein--Hawking entropy/area law\cite{bekhawk}\ of Black Hole
Thermodynamics can be derived as a truly statistical result.

This was first done for five dimensional Riessner--Nordstr\"om black
holes\cite{andycumrun}, and later shown  for the four dimensional
case\cite{us,them}.

The basic idea is simple\cite{andycumrun}. The black holes of interest
were embedded into (say) $K3{\times}T^2$ compactified type IIB string
theory as a BPS state.  The abelian (Maxwell) fields they carry were
embedded into the R-R sector of the compactified theory.  The final
embedding resembles a macroscopic string in six dimensions with a
certain amount of R-R charge, together with a NS-NS charge
representing momentum in one of the $T^2$ directions.

So far this is just a black hole dressed up with a stringy
embedding. The area of its horizon can be computed, as normal, and
thus the entropy may be computed using the Bekenstein--Hawking law. We
can also compute all of the R-R and NS-NS charges it carries using
ordinary flux integrals. The important point comes when we try to
evaluate the entropy as a statistical quantity.

The internal structure of the black hole is a region we can only hope
to probe with strong coupling string theory/quantum gravity.  However,
we can deform the theory to weak coupling (as the black hole is a BPS
state of the string theory) and compute the entropy there.  At weak
coupling, the unique BPS saturated object in the string theory with
the same quantum numbers as the black hole we saw at strong coupling,
is a bound state of certain types of D--branes.

We can compute precisely how many D--branes are present in the
composite due to the fact that we know the R-R charge of the
composite, and the fact that D--branes carry the basic units of R-R
charge\cite{gojoe}.

The NS-NS charge corresponding to momentum is identified with the
total momentum distributed among the fundamental strings connecting
the constituent D--branes of  the composite.  Evaluating the
number of ways to distribute this quantum number amongst the bound
state's constituents gives precisely the entropy.

\section{Application III: Lower Dimensional ${\cal N}{=}1$ String Vacua} 

The basic orientifold example of the type I string can be extended to
more complicated situations\cite{petrii,openstuff}.  Developing the
orientifold technology further, one can study a much larger class of
string models than was previously possible with more traditional
techniques. (The technology of constructing consistent string models in
lower dimensions was largely limited to the closed string theories, as
the unoriented and open string sectors were not as easily manipulated.)

The orientifold group is simply the group of symmetries of the string
theory which we might like to consider gauging, in a way anologous to
what was carried out for $\Omega$ in the ten dimensional example of
section~3.

We consider discrete symmetries, which can be either purely of
world--sheet ($\Omega$) or of spacetime (like those used to form a
conventional orbifold).  Perturbative consistency of the models
obtained by gauging such a group of symmetries is ensured
by\cite{ericjoe}:

\begin{itemize}
\item{Finding  a faithful representation of the symmetries of the
orientifold group on the open string sectors (D--Branes).}
\item{Insisting on  tadpole cancellation at one--loop.}
\end{itemize}

In general, there will be tadpole cancellation associated with all the
species of R-R $(p{+}1)$--forms which can appear in the model,
requiring the presence of a D$p$--brane. Tadpole cancellation is
another way of seeing that the R-R $(p{+}1)$--form's field equation is
satisfied.

In this way, one can study large families of ${\cal N}{=}1$ models in
lower dimensions with interesting and important
properties\cite{petrii,openstuff,ericjoe,atishi,ericme,atishii}.

Constraints of time and space do not permit me to describe them further here.

\bigskip

\noindent{\bf Acknowledgements:} My understanding of many of the issues
which I presented above has been sharpened in numerous discussions
with Rob Myers and Joe Polchinski over the last six months. I would
like to thank Samantha Butler for helpful comments on the manuscript.

\medskip

\centerline{
\epsfxsize=1.7in\epsfbox{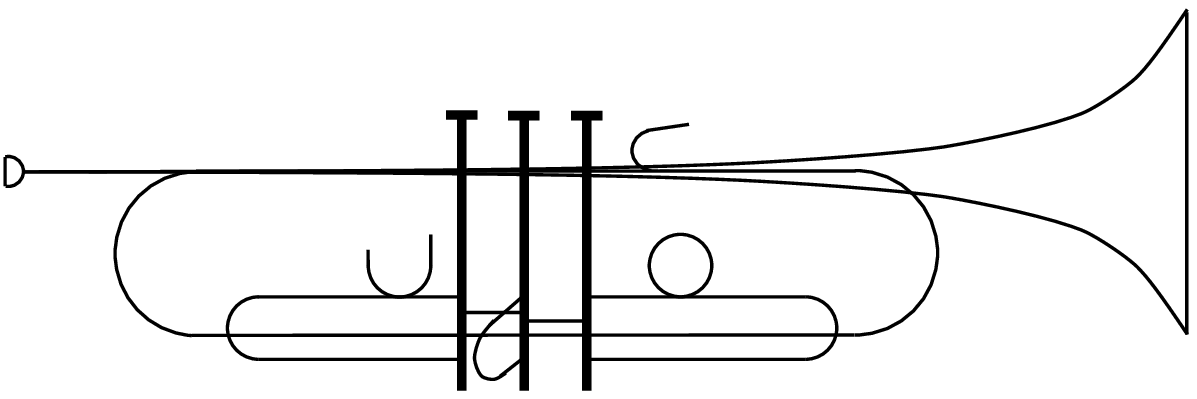}
}

\end{document}